\documentclass[preprint,12pt]{elsarticle}
\usepackage{nicefrac}
\usepackage{algorithm}
\usepackage{algorithmic}
\usepackage{amssymb}
\usepackage{amsthm}
\usepackage{amsmath}
\usepackage{graphicx}
\usepackage{hyperref}
\usepackage{xcolor}
\journal{Computer Physics Communications}

\newcommand{\beq}{\begin{equation}}
\newcommand{\br}{{\mathbf r}}
\newcommand{\eeq}{\end{equation}}
\newcommand{\bea}{\begin{eqnarray}}
\newcommand{\eea}{\end{eqnarray}}
\newcommand{\balg}{\begin{align}}
\newcommand{\ealg}{\end{align}}

\newcommand{\eqn}[1]{\mbox{Eq.\hspace{1pt}(\ref{#1})}}
\newcommand{\eqs}[2]{\mbox{Eq.\hspace{1pt}(\ref{#1}--\ref{#2})}}

\def\brp{{\mathbf{r}^{\prime}}}

\def\br{{\mathbf{r}}}

\def\vdw{{van der Waals}}

\def\F{{\mathcal{F}}}

\begin{document}
\begin{frontmatter}
\title{eQE 2.0: Subsystem DFT Beyond GGA Functionals}
\author[R_chem]{Wenhui Mi \corref{ca}}
\ead{wenhui.mi@rutgers.edu}
\author[R_chem]{Xuecheng Shao \corref{ca}}
\ead{xuecheng.shao@rutgers.edu}
\author[R_chem,kitware]{Alessandro Genova \corref{ca}}
\ead{alessandro.genova@kitware.com}
\author[milano]{Davide Ceresoli \corref{ca}}
\ead{davide.ceresoli@cnr.it}
\author[R_chem,R_phys]{Michele Pavanello \corref{ca}}
\ead{m.pavanello@rutgers.edu}
\cortext[ca]{Corresponding authors}
\address[R_chem]{Department of Chemistry, Rutgers University, Newark, NJ 07012, USA}
\address[R_phys]{Department of Physics, Rutgers University, Newark, NJ 07012, USA}
\address[kitware]{Kitware Inc., 1712 U.S. 9 Suite 300, Clifton Park, NY 12065, USA}
\address[milano]{CNR-SCITEC c/o Dipartimento di Chimica, Universit{\`a} degli studi di Milano via Golgi 19, 20133 Milano, Italy}
\begin{abstract} 
	By adopting a divide-and-conquer strategy, subsystem-DFT (sDFT) can dramatically reduce the computational cost of large-scale electronic structure calculations. The key ingredients of sDFT are the nonadditive kinetic energy and exchange-correlation functionals which dominate it's accuracy.  Even though, semilocal nonadditive functionals find a broad range of applications, their accuracy is somewhat limited especially for those systems where achieving balance between exchange-correlation interactions on one side and nonadditive kinetic energy on the other is crucial. In eQE 2.0, we improve dramatically the accuracy of sDFT simulations by (1) implementing nonlocal nonadditive kinetic energy functionals based on the LMGP family of functionals; (2) adapting Quantum ESPRESSO's implementation of rVV10 and vdW-DF nonlocal exchange-correlation functionals to be employed in sDFT simulations; (3) implementing ``deorbitalized'' meta GGA functionals (e.g., SCAN-L).  We carefully assess the performance of the newly implemented tools on the S22-5 test set. eQE 2.0 delivers excellent interaction energies compared to conventional Kohn-Sham DFT and CCSD(T). The improved performance does not come at a loss of computational efficiency. We show that eQE 2.0 with nonlocal nonadditive functionals retains the same linear scaling behavior achieved in eQE 1.0 with semilocal nonadditive functionals.

%	Benchmarks assessing the performance of the new implementations indicate a strong improvement of the accuracy compared to typical semilocal functionals while still maintaining a linear scaling behavior with system size. Thus, eQE 2.0 is expected to considerably extend the applicability of sDFT for materials science, {physics,} and chemistry applications.
%In the first release of eQE, the parallelization scheme and convergence approaches. In this version, eQE departs form semilocal nonadditive functionals (for both XC and NAKE). 
\end{abstract}

\end{frontmatter}
{\bf \noindent PROGRAM SUMMARY\\}
\begin{small}
\noindent
{\em Program title: eQE\\}
{\em Licensing provisions: GNU\\}
{\em Distribution format: .tar.gz, git repository\\}
{\em Developer's repository link: https://gitlab.com/Pavanello/eqe\\}
{\em Programming language: Fortran 90 \\}
{\em External routines/libraries: BLACS, MPI \\}
{\em Operating system: Unix-like (Linux, macOS, Windows Subsystem for Linux)\\}
{\em Nature of problem: Solving the electronic structure of molecules and materials with subsystem density functional theory\\}
{\em Solution method: eQE \\} 
{\em Additional comments: url of stable release http://eqe.rutgers.edu } 
\end{small}

\section{Introduction}
\label{sec:intro}
% KS-DFT is most wildely used and limited
Kohn--Sham density functional theory (KS-DFT) \cite{hohe1964,kohn1965} is the most widely used approach for {\it ab initio} electronic structure simulations and has been employed for scientific discoveries across a broad range of disciplines such as chemistry, physics, and materials science. The success of KS-DFT is due to the good balance between computational cost and accuracy. However, due to its cubic scaling with respect to the number of electrons, $N_{e}$, KS-DFT calculations are limited in the system sizes they can approach.
% the possiable approches limitations and sDFT's advantage

Several alternative approaches have been proposed to access system sizes that relate to experimentally relevant microstates. Among them, we recall linear-scaling KS-DFT \cite{goedecker1999linear,bowl2012,Liou_2020,Sena_2011}, orbital-free DFT (OF-DFT) \cite{wang2000,witt2018orbital,ATLAS,shao2018large,huan2010,mi2019LMGP,luo2018simple,karasiev2012issues}, subsystem-DFT (sDFT) \cite{Mi_2019,weso2015,gome2012,jaco2014,krish2015a,nafz2014,LDA_NAKE,weso1993}, a combination of the two \cite{Andermatt_2016} and others \cite{porezag1995construction,elstner1998self}. These approaches can achieve
(quasi)-linear scaling by exploiting specific features, such as the  ``nearsightedness" of the
electronic structure which arises in many non-conductive systems \cite{goedecker1999linear,Liou_2020}.
%Moreover,
%due to the large prefactor of the linear-scaling, the computational
%efficiency advantages occur need to simulate systems comprised of up to
%thousands of atoms\cite{Moussa_2019}. 
%In contrast, the total energy functional in OF-DFT (including the noninteracting kinetic energy) is strictly a pure functional of the electron density whose computational complexity is low by construction. However, even though OF-DFT is certainly a promising method \cite{huan2010,mi2019LMGP,luo2018simple}, the use of density functionals to approximate the non-interacting kinetic energy can result in approximations detrimental to the accuracy of the simulations\cite{witt2018orbital,karasiev2012issues}.  
  
% sDFT is a powerfull tool for large-scale simutions
%The accuracy is determibed by the accuracy of the functionals

sDFT adopts a divide-and-conquer strategy, whereby a system is split into smaller but interacting subsystems. In this way, the cubic scaling of conventional KS-DFT can be reduced. Linear scalability in sDFT is achieved by modeling the interactions between subsytems with pure density functionals: the nonadditive exchange-correlation (NAXC), the nonadditive kinetic energy (NAKE), and the long-range Coulomb interaction between the subsystems' charge densities. Among them, the only available exact functional is the  Coulomb interaction while the other two need to be approximated. Thus, the accuracy of sDFT is dictated by NAKE and NAXC.

% Overview Previous work of eQE-- GGA level functionals
In this work, we present version 2.0 of embedded Quantum ESPRESSO (eQE) \cite{fderelease}, a code that implements sDFT based on Quantum ESPRESSO \cite{qe,giannozzi2020quantum,qe_new}. In the first release of eQE \cite{fderelease}, we successfully implemented semilocal (GGA) level nonadditive functionals for both NAXC and NAKE with the following main features: (1) a scheme of parallel execution to distribute the workload across subsystems resulting in low data communication achieving high parallel efficiency, (2)  {\it ab initio} molecular dynamics (AIMD), and (3) applicability to periodic systems. Many large systems currently outside KS-DFT's realm of applicability have been successfully studied by eQE \cite{Umerbekova_2020,Umerbekova_2018,Kumar_2017,Mi_2019a,Genova_2016a,genova2014,geno2015a}.  

eQE 2.0 is capable of deploying nonlocal and meta-GGA (mGGA) XC and NAKE
functionals yielding highly accurate simulations of weakly interacting
subsystems. It is
well known that, due to their dependence on the density and its derivative in
only one point, semilocal XC functionals inherently lack the ability to capture
intermediate-range and long-range dispersion interactions \cite{misq2003,pern2009b,klim2010,klim2012,vydr2011,lang2005,dobs2012,Ferri_2015,Hermann_2017}. However,  dispersion
interactions play a crucial role in systems that are weakly bound and amenable
to be treated by sDFT. mGGA and nonlocal XC functionals address these issues
as mGGAs capture intermediate-range correlations due to their dependence on
higher order derivatives of the density beyond the gradient or the kinetic energy density. 
Nonlocal
functionals capture long-range dispersion interactions because they encode a
dependency on the density on more than a single point at a time. 
%Particularly,
%the SCAN metaGGA functional\cite{Sun_2015} and the rVV10 nonlocal
%functional\cite{rvv10} are known to perform well.  

As mentioned, GGA NAKEs are widely used \cite{jaco2014,krish2015a,weso2015}. However, due to their inability to describe the inherent non-locality of the kinetic energy, GGAs work well only when the inter-subsystem density overlap is weak. In principle, nonlocal NAKEs have the potential to obtain more reliable results \cite{witt2018orbital,Mi_2019} than GGAs. In practice, however, most of the available nonlocal functionals are not suitable to be used as NAKEs because they are optimized for bulk systems where the electron density is nonzero everywhere. Conversely, the subsystem densities in sDFT feature at least one non-periodic dimension where they decay to zero. 

Recently, we have developed a new generation of nonlocal
 kinetic energy functionals \cite{mi2018nonlocal,mi2019LMGP} and further adopted them as NAKEs \cite{Mi_2019} considerably improving the accuracy of sDFT in terms
of both energy and electron density. However, the original implementation \cite{Mi_2019} is computationally very expensive compared to typical GGA NAKEs. In eQE 2.0, we implement a completely new scheme to evaluate nonlocal NAKEs that brings down the computational cost to almost GGA-level without sacrificing accuracy.

%In this work, we successfully released new version of eQE beyond GGA functionals. These mainly include:  (1) propose a new efficient scheme to implement the LMGP nonlocal NAKEs. (2) Implemented rVV10 nonlocal XC functional with an efficient approach. (3) Implemented deorbitalized mGGA XC functional (SCAN-L) in sDFT as typical GGA XC functional scheme.  

The remainder of this paper is organized as follows. A brief review of sDFT and the previous version of eQE is given in section \ref{sec2}. Details of the implementation of the new functionals and benchmark calculations are discussed in section \ref{sec:new}. Finally, in section \ref{sec:conclusion} we summarize this work and conclude with several potential future developments aimed to further increase the range of applicability of sDFT.
 \section{Brief review of sDFT and eQE }
\label{sec2}
% formulations of eQE
In sDFT, the system of interest is split into interacting subsystems of smaller size. Accordingly, the total electron density, $n(\br)$, is partitioned into subsystem contributions:  
\beq
n(\mathbf{r}) =\sum_{I}^{N_{s}} n_{I} (\br), 
\eeq
where $N_{s}$ is the total number of subsystems and $n_{I}(\br)$ is the electron density of the subsystem $I$. In this way, the total energy functional of the system, $E_{\rm sDFT}[\{n_I\}]$, contains additive and nonadditive contributions encoding intra- and inter-subsystem interactions. Namely,
\beq
 E_{\rm sDFT}[\{n_I\}] = \sum_{I}^{N_{s}}E_{KS}[n_{I},v_{ext}^{I}] + T_s^{\rm nadd}[\{n_{I}\}] + E_{\rm xc}^{\rm nadd}[\{n_{I}\}]+E_{\rm Coul}^{\rm nadd}[\{n_{I}\},\{v_{ext}^{I}\}]
\eeq 
where $v_{ext}^{I}$ is the external potential associated with subsystem $I$. The energy functional for each subsystem, $E_{KS}[n_{I},v_{ext}^{I}]$, is the same as the energy functional of typical KS-DFT with $v_{ext}^{I}$ as external potential and $n_I$ as electron density. $T_{s}^{\rm nadd}$, $E_{\rm xc}^{\rm nadd}$, and $E_{\rm Coul}^{\rm nadd}$ are the NAKE and NAXC energy functionals, and the nonadditive Coulomb energy, respectively. All these nonadditive energy functionals share the same definition:
\beq
F^{\rm nadd} [\{n_{I}\}]= F[n] -\sum_{I}^{N_{s}} F[n_{I}].
\eeq
Among them only $E_{\rm Coul}^{\rm nadd}$ can be expressed exactly, the others need to be approximated.   

Variational minimization of $E_{\rm sDFT}$ with respect to variations of the electron density of each subsystem leads to the following coupled KS-like equations,
\beq
\left[-\frac{1}{2}\nabla^2 + v_{\rm KS}^{I}(\br)  +  v^{I}_{\rm emb}(\br) \right] \phi_{i}^{I}(\br) = \varepsilon^I_{i} \phi_{i}^{I} (\br),
\eeq 
where $v_{\rm KS}^{I}$, $v_{\rm emb}^{I}$, $\phi_{i}^{I}$ are the KS--potential of the isolated subsystem, the embedding potential containing the functional derivatives of the nonadditive energy terms with respect to $n_I$, and the orbitals of subsystem $I$, respectively. The embedding potential can be written as follows:
\begin{equation}
 \label{embpot}
 \upsilon^{I}_{emb}(\br)=\sum^{N_s}_{J\neq I}\left[\int \frac{n_J(\brp)}{|\br-\brp|}d\brp+v_{ext}^J(\br)\right]+\frac{\delta T_{\rm s}^\text{nad}[\{n_{I}\}]}{\delta n_{I}(\br)}+\frac{\delta E^\text{nad}_{\rm xc}[\{n_{I}\}]}{\delta n_{I}(\br)}.
\end{equation} 

The original version of eQE, implemented sDFT in the following way:
\begin{enumerate}
	\item The construction of subsystem Hamiltonians and their diagonalizations are performed independently for each subsystem in parallel (all subsystems at the same time) and in a reduced subsystem-centered simulation cell.
	\item Subsystems can be assigned a custom number of processors to optimize load balance.
	\item The parallelization scheme is hierarchical designed to reduce inter-node communication. 
	\item An additional global DIIS layer \cite{fderelease,pula1982} to reduce the number of SCF cycles.
	%\item Subsystem‐specific basis sets, lattice vectors, and integration grids to allow the efficiently evaluation of density, potentials, and total energy.
 \end{enumerate}  
%\begin{figure}
%\includegraphics[width=1.0\textwidth]{./fig/Efficiency.pdf}
%\caption{\label{efficiency} Scaling of eQE when simulating systems composed of 32, 64, 128, and 256 CO$_2$ molecules maintaining a constant density. The corresponding numbers of MPI tasks are 64, 128, 256, 512. The dotted line (black) represents the ideal linear scaling of the algorithm with 100\% parallel efficiency. In blue we report the performance of the KS-DFT code (QE) for the same systems.}
%\end{figure}

By adopting these steps, eQE achieved high parallel efficiency as showcased in several application works \cite{Umerbekova_2020,Kumar_2017,Mi_2019a,Genova_2016a}. 
%Here, we show another case that how the eQE software is able to maintain the low-scaling computational cost when both the number of processors and the size of the system. Figure. \ref{efficiency} shows the wall timings of eQE and the latest version of Quantum ESPRESSO\cite{qe} in simulating systems composed of 32 to 256 CO$_2$ molecules with a constant density and corresponding processors for each molecule. Obviously, it indicates that the performance of eQE can be approach to ideal linear scaling and parallel efficiency in simulating proper systems with reasonable settings. 
We need to stress that the computational efficiency and accuracy of eQE is system dependent. In general, the larger the number of subsystems and the smaller their size, the better eQE performs compared to KS-DFT codes.      
 
% The key nonadditive part

% The way of evaluation of the Nonadditive 
\section{New Features and upgrades in eQE 2.0}
\label{sec:new}
In this section, We discuss the main features and updates of eQE since its original release in 2017 \cite{fderelease}. 

\subsection{Nonlocal NAKEs}
Recently, we developed fully nonlocal sDFT \cite{Mi_2019}, which considerably improves on the accuracy of semilocal sDFT. However, in the original implementation, the computational cost was orders of magnitude larger than semilocal NAKEs. In this work, the issue was resolved by presenting a new implementation of nonlocal NAKE which achieves the same accuracy in the results at essentially the cost of semilocal functionals. 

%For the sake of simplicity, GGA functionals, $F[n,\nabla n]$, the corresponding potential $v[n,\nabla n]$, and energy density $\tau[n,\nabla n]$ are written as $F[n]$, $v[n]$, and $\tau[n]$, respectively.
The typical form of a nonlocal kinetic energy functional is: 
\beq
T_s[n] =\underbrace{T_{TF}[n] + T_{vW}[n]}_{T_{TV}[n]} + T_{NL}[n]
\eeq
where, $T_{TF}[n]$, $T_{vW}[n]$, and $T_{NL}[n]$  are the Thomas-Fermi (TF) \cite{fermi1927,thom1927} and von Weizs\"acker (vW) \cite{weiz1935} functionals, and lastly the nonlocal part of the functional, respectively.  The nonlocal part is in practice a two-point functional defined by a double integration of the electron density at two different points in space. The interaction between two points is described by the so-called kernel, $\omega$:
\beq
T_{NL}[n]=\int \int n^{\alpha}(\br)\omega[n](\br,\br')n^{\beta}(\br')d\br d\br'
\eeq 
where $\alpha$ and $\beta$ are positive numbers. 
The corresponding potential is obtained by functional derivative with respect to the density:
\beq
v_{T_{s}}(\br)=\frac{\delta T_{TV}[n]}{\delta n(\br)} +\frac{\delta T_{NL}[n]}{\delta n(\br)}=v_{TV} (\br)+ v_{NL}(\br).
\eeq
As discussed in our previous work \cite{Mi_2019}, direct implementation of this formalism for calculating the kinetic potential leads to numerical instabilities for both terms in the region of low electron density. In this region, the von Weizs\"acker potential is inaccurate due to its dependence on the Laplacian of the electron density which is noisy for low density. The nonlocal part shares a similar issue. In eQE 2.0, we resolve these issues as follows. 

% For STV
For the $T_{TV}[n]=T_{TF}[n] + T_{vW}[n]$ term, the corresponding GGA formalism 
reads,
\beq
T_{TV}[n]=\int \tau_{TV}(\br) d\br\int \tau_{TF}(\br) F_{TV}[s](\br) d\br,
\eeq
where $s$ is the  dimensionless reduced density gradient,  $s=\frac{1}{2(3\pi^{2})^{1/3}}\frac{|\nabla n(\br)|}{n^{3/4}(\br)}$, the enhancement  factor $F_{TV}(s)=1+\frac{5}{3}s^{2}$, and $\tau_{TF}(\br) = \frac{3}{10}(3\pi^2)^{\frac{2}{3}}n^{\frac{5}{3}}(\br)$. To eliminate the numerical inaccuracies arising at large $s$, 
we developed a numerically stable enhancement factor 
\beq
\label{stv}
F_{STV}(s)=1.0 + \frac{5}{3}\left(\frac{s^{2}}{1.0+as^{2}}\right),
\eeq
which is used in place of the original $F_{TV}[n]$, with $a=0.01$.

% For mix potentials scheme
 We choose the LMGP family of functionals \cite{Mi_2019} as the nonlocal kinetic energy functional in eQE 2.0. LMGP's potential term, $v_{NL}(\br)$, can be written as follows:
\beq
v_{NL}(\br)=n^{-\nicefrac{1}{6}}(\br)\F^{-1}\bigg[\widetilde{n^{\nicefrac{5}{6}}}(\mathbf{q})\omega[n(\br)](q)\bigg](\br),
\label{vnl}
\eeq
where $\widetilde{n^{\nicefrac{5}{6}}}(\mathbf{q})=\F\big[n^{\nicefrac{5}{6}}(\br)\big](\mathbf{q})$, $\omega[n(\br)](q)$ is the nonlocal kernel expressed in reciprocal space,  $\F$ and $\F^{-1}$ represent forward and inverse Fourier transforms, respectively. There are two important details to note:
\begin{enumerate}
	\item the nonlocal kinetic potential has a $n^{-\nicefrac{1}{6}}(\br)$ factor which can lead to numerical noise in the low electron density regions. 
	\item The kernel of LMGP is density-dependent and of spherical symmetry, e.g., $\omega[n](\br,\brp)=\omega[n(\br)](|\br-\brp|)$.
\end{enumerate}
To eliminate issues related to point (1) above, a density weighted mix of GGA and nonlocal potentials was implemented \cite{Mi_2019}. Namely,
\beq
v_{T_{s}}[n](\mathbf{r})=\left(v_{N L}[n](\mathbf{r})+v_{S T V}[n](\mathbf{r})\right) W[n](\mathbf{r}) 
+v_{G G A}[n](\mathbf{r})(1-W[n](\mathbf{r}))
\eeq
where $W[n](\br) = \frac{n(\br)}{n_{max}}$, $n_{max}$ is the maximum value of electron density in the system and $v_{GGA}$ is the potential of a well-established GGA kinetic energy functional (we use revAPBEK \cite{revAPBEk}). The kinetic energy can be obtained by line integration \cite{chai2007,Mi_2019}:
\beq
T_{s}[n]=\int n(\mathbf{r}) d \mathbf{r} \int_{0}^{1} v_{T_{s}}\left[n_{t}\right](\mathbf{r}) d t,
\label{Tint}
\eeq  
where $n_t(\mathbf{r})=tn(\mathbf{r})$. This required using a set of 40 or more $t$ points to obtain the converged results and with that 40 or more kinetic potentials $\{v_{T_s}[n_t](\br)\}$ needed to be evaluated. 

To reduce the computational cost and still maintain numerical stability and accuracy, eQE 2.0 features a kinetic energy density mix where the total KEDF can be approximated as:
\beq
\begin{aligned}
\label{T:MKE}
T_{s}[n]&=\int \tau[n](\br) d\br \\
&=\int W[n](\br)\bigg[\tau_{NL} [n](\mathbf{r}) + \tau_{STV}[n](\br)\bigg] + \bigg(1-W[n](\br)\bigg)\tau_{GGA}[n](\br ) d\br
\end{aligned}
\eeq
where $\tau[n]$, $\tau_{STV}[n]$, $\tau_{GGA}[n]$, $\tau_{NL}[n]$ are the kinetic energy density for total KEDF, $T_{STV}[n]$ and $T_{GGA}[n]$ (revAPBEk \cite{revAPBEk}), respectively. 
The corresponding kinetic potential is
\beq
\begin{aligned}
\label{V:MKE}
v_{s}[n](\br)& =\frac{\delta T_{s}[n]}{\delta n(\br)}=\frac{\partial W[n]}{\partial n(\br)}\bigg[ \tau_{NL}[n](\br)  + \tau_{STV}[n](\br)-\tau_{GGA}[n](\br)\bigg]\\
	&+W[n](\br)\bigg[ v_{NL}[n](\br)  +v_{STV}[n](\br) -v_{GGA}[n] (\br) \bigg] + v_{GGA}[n](\br).
\end{aligned}
\eeq
where $
\frac{\partial W[n]}{\partial n(\br)} =\frac{1}{n_{max}}, \,v_{GGA}[n](\br) =\frac{\partial \tau_{GGA}[n]}{\partial n(\br)} -\nabla \cdot \frac{\partial \tau_{GGA}[n]}{\partial \nabla n(\br)}, \, v_{STV}[n](\br)= \frac{\partial \tau_{STV}[n]}{\partial n(\br)} -\nabla \cdot \frac{\partial \tau_{STV}[n]}{\partial \nabla n(\br)}$

In practice, the above equations for both KEDF \eqn{T:MKE} and kinetic potential \eqn{V:MKE} can be further separated into three terms. Namely,
\begin{align}
	T_{s}[n] & = E_{STV}[n] + E_{GGA}[n] + E_{NL}[n],\\
	v_s[n](\br) & =V_{STV}[n](\br) + V_{GGA}[n](\br) + V_{NL}[n](\br)
\end{align}
and each part can be evaluated in the following way: 
\begin{enumerate}
\item STV term
\begin{align}
	V_{STV}[n](\br) &= W[n](\br)v_{STV}[n](\br) + \frac{\tau_{STV}[n](\br)}{n_{max}},\\
	E_{STV}[n] &=\int  W[n](\br)\tau_{STV}[n](\br) d\br.
\end{align}
\item GGA term
\begin{align}
	V_{GGA}[n](\br) &= \left[1-W[n](\br)\right]v_{GGA}[n](\br) -\frac{\tau_{GGA}[n](\br)}{n_{max}},\\
	E_{GGA}[n] &=\int \left[1-W[n](\br)\right]\tau_{GGA}[n](\br)d \br.
\end{align}
\item NL term
\beq
V_{NL}[n](\br) =W[n](\br)v_{NL}[n](\br) +\frac{\tau_{NL}[n](\br)}{n_{max}}
\eeq
where the nonlocal kinetic density $\tau_{NL}$ needs to be approximated because the functional integration procedure in \eqn{Tint} is too expensive. From \eqn{vnl} notice that $v_{NL}[n]\sim n^{2/3}$, with $W[n]\propto n$, $\tau_{NL}$ can be approximated as $\tau_{NL}[n](\br) \simeq\frac{3}{8} n(\br) v_{NL}[n](\br)$. Thus, the nonlocal kinetic energy can be approximated as:  
\beq
E_{NL}[n] =\int \tau_{NL}[n](\br) W[n](\br) d\br . 
\eeq
\end{enumerate}
 In this way, each time the kinetic energy/potential is evaluated, only two GGA KEDFs and one fully nonlocal term need to be calculated, dramatically reducing the computational cost compared to the original eQE implementation.
\begin{figure}
\label{Accuracy} 
\includegraphics[width=1.0\textwidth]{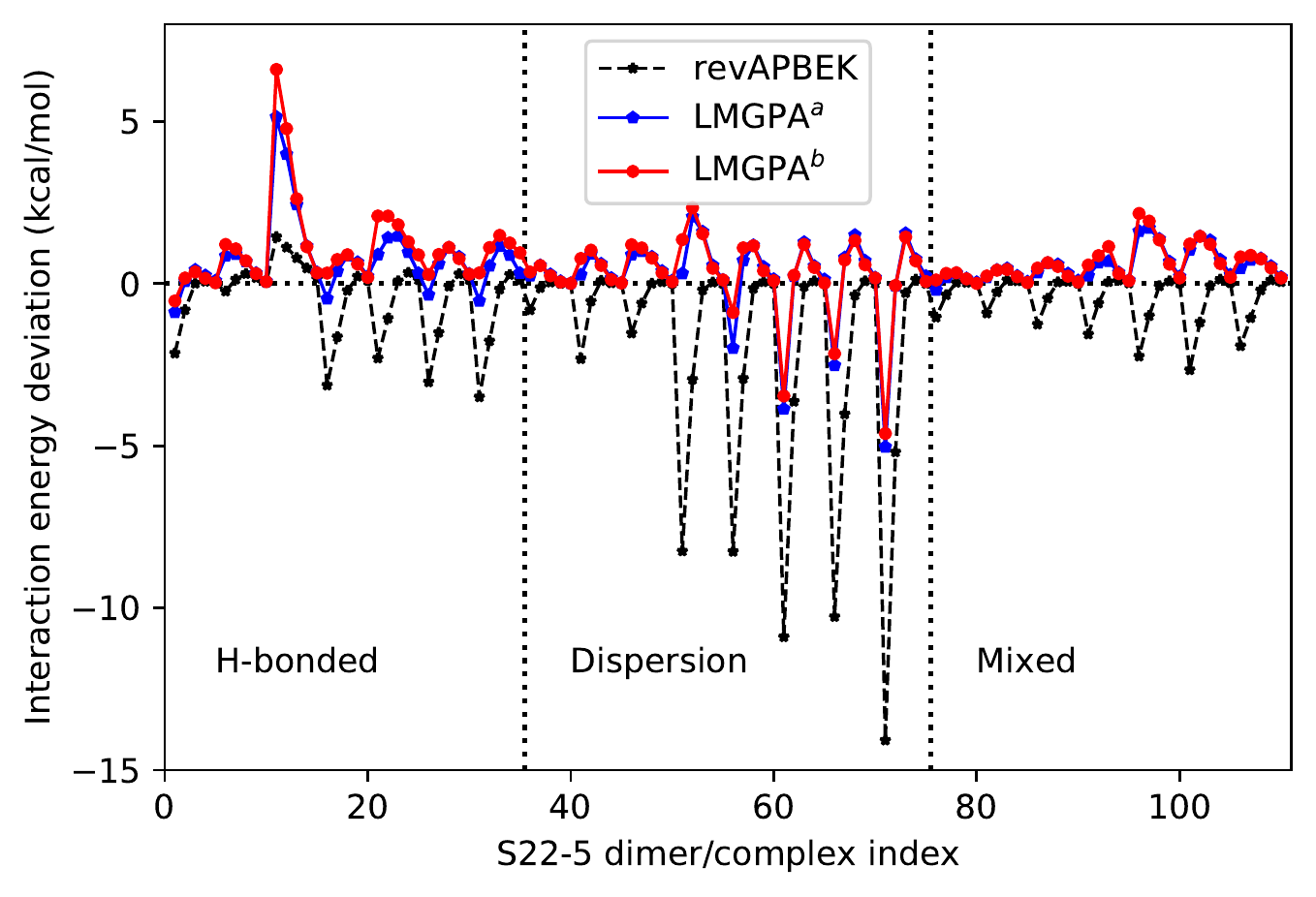}
	\caption{Interaction energy deviations against the corresponding KS-DFT results for the S22-5 test set obtained by eQE with the following NAKEs: revAPBEK (GGA), LMGPA$^a$ (original, nonlocal), and LMGPA$^b$ (this work, nonlocal). LMGPA is the LMGP functional with ``arithmetically symmetrized'' kernel as explained in \cite{mi2018nonlocal}. PBE XC functional is adopted in all calculations. The indices on the $x$-axis correspond to complexes listed in Table S1 of the Supporting Information from Ref.\,\cite{Mi_2019}}
\end{figure}

The ability of nonlocal sDFT to predict interaction energies and electron densities for the S22-5 test set (noncovalently interacting complexes at
equilibrium and displaced geometries\cite{Grafova_2010}) has been demonstrated in our previous work \cite{Mi_2019}.  To quantify the accuracy of the new implementation, we select the same test set with the same calculation settings as before. The interaction energy deviations between sDFT with different NAKEs, such as revAPBEK and LMGPA (i.e., LMGP with arithmetically symmetrized kernel \cite{mi2018nonlocal} which we implemented in the original and new version of eQE) against the corresponding KS-DFT reference are shown in Figure \ref{Accuracy}. 

The reason why we compare sDFT with KS-DFT is that in KS-DFT the total noninteracting kinetic energy is exact, while in sDFT the nonadditive part of the noninteracting kinetic energy is approximate. If the two methods deliver similar results for a given choice of exchange-correlation (we use PBE for this comparison) then it means that the approximate nonadditive functional used in sDFT is accurate. Later we will also compare against CCSD(T) benchmark values. As expected, the LMGPA nonlocal NAKE considerably improves on the results of semilocal sDFT, especially when the dimer distances are shorter than the equilibrium distance (i.e., for the S22-5(0.9) subset). Most importantly, the results from the new version of LMGPA are almost on top of the original version, showing that the new implementation maintains an excellent level of accuracy but is computationally much cheaper. A similar result is obtained when a different XC functional is used (such as a nonlocal XC functional) \cite{Mi_2019}.

Hereafter, the new implementation of LMGPA (noted as LMGPA$^b$ in the figure captions) is simply referred to as LMGPA. 

%In eQE, once the nonlocal KEDF in hand, $T_{s}[n]$, $T_{s}[n_{I}]$ and the corresponding kinetic potentials can be evaluated in both the whole system and the subsystems. The evaluation of  kernels and corresponding potentials and energies can be independently performed within the corresponding cells. The cell-independent kernel function $\omega(\eta)$ (where $\eta=|\mathbf{q}|/2k_{F}$, $\mathbf{q}$ is the reciprocal space variable for $|\br -\brp|$ and $k_{F}$ is the Fermi vector) can be directly read from the kernel file. Thus, for this kind of nonlocal NAKEs, the implementations are exactly the same, except for the kernel files. 

\subsection{Nonlocal NAXCs}
\begin{figure}[htp]
\includegraphics[width=0.8 \textwidth]{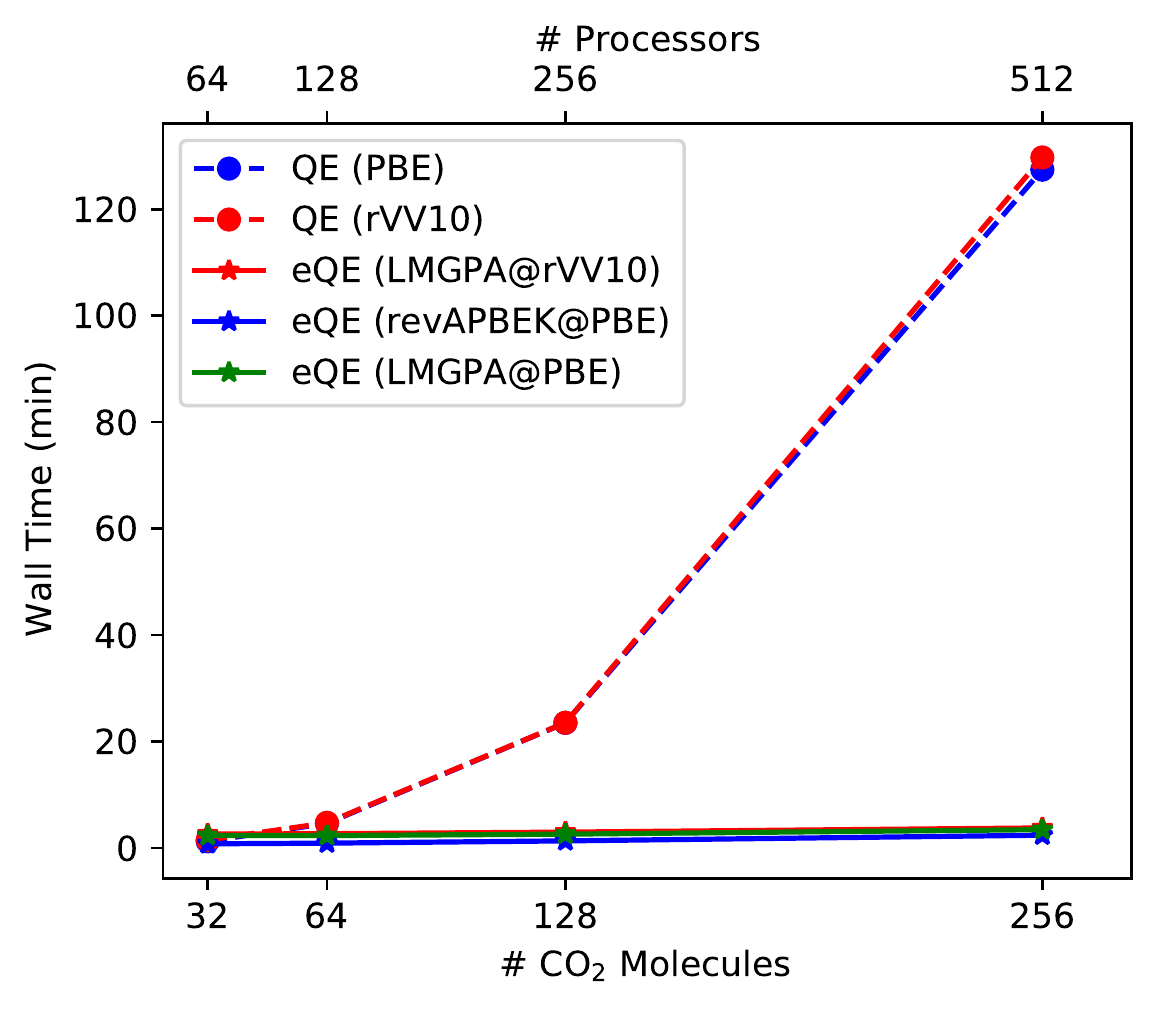}
	\caption{\label{FNL} Wall times when simulating supercells containing 32, 64, 128, and 256 CO$_2$ molecules. The corresponding numbers of MPI tasks are 64, 128, 256, 512. Full lines are eQE simulations with different NAKE funtionals. Doted lines are KS-DFT (QE version 6.6) simulations for the same systems and same cutoff for the plane waves (i.e., 40 Ry for the wavefunctions and 400 Ry for the electron density).}
\end{figure}
It is well known that semilocal XC functionals struggle to describe long-range correlation effects such as dispersion interactions. However, dispersion interactions play a critically important role in a wide variety of materials, especially those composed of weakly interacting subsystems. Thus, it is necessary to adopt NAXC functionals that go beyond GGA to systematically improve the performance of sDFT.    

\begin{figure}
\label{s22bench} 
\includegraphics[width=1.0\textwidth]{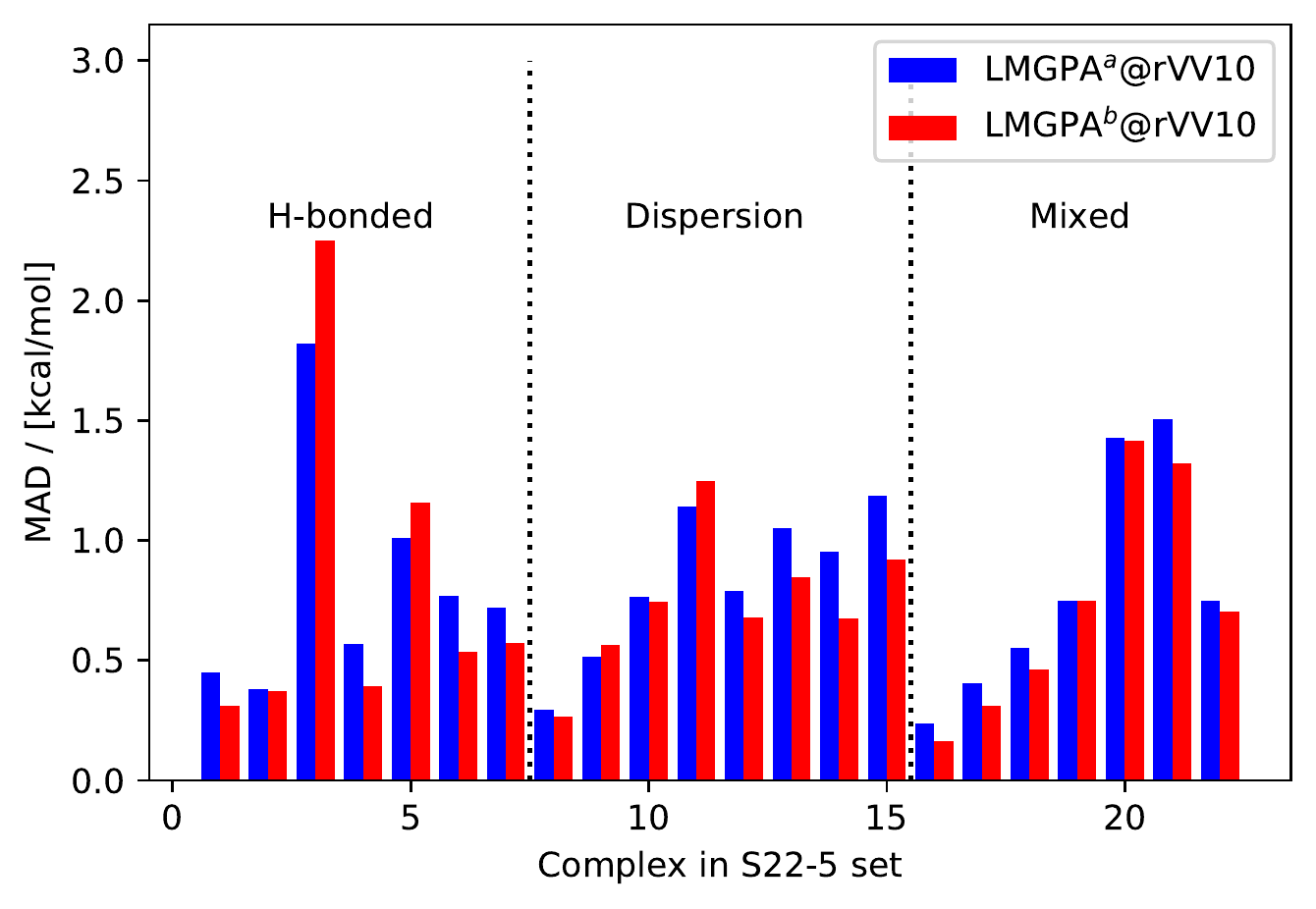}
	\caption{Interaction energy deviations against CCSD(T) benchmark values for the S22-5 test set. sDFT calculations carried out with eQE with the following NAKEs: revAPBEK (GGA), LMGPA$^a$ (original, nonlocal), and LMGPA$^b$ (this work, nonlocal). LMGPA is the LMGP functional with ``arithmetically symmetrized'' kernel as explained in \cite{mi2018nonlocal}. rVV10 XC functional is adopted in all calculations. The indices on the $x$-axis correspond to complexes listed in Table S1 of the Supporting Information from Ref.\,\cite{Mi_2019}}
\end{figure}

Many nonlocal XC functionals have been proposed \cite{dion2004van,lee2010higher,vydrov2009nonlocal,rvv10,vydrov2010nonlocal}. Among them, rVV10 \cite{rvv10} stands out for its accuracy across not only non-covalently bound complexes but also covalent, ionic, and metallic solids. Most importantly, the computational cost can remain low \cite{roma2009}.  Thus, we adapted the existing rVV10 implementation of Quantum ESPRESSO \cite{rvv10} for eQE, where an evaluation of the functional and corresponding potential in the supersystem (physical cell) and in the subsystem-centered cells are needed. The resulting fully nonlocal sDFT compares favorably against CCSD(T) benchmark values for the interaction energies of the S22-5 set as can be seen in Figure \ref{s22bench}.

To quantify the computational efficiency and scaling, we simulated supercells of 32, 64, 128, and 256 CO$_2$ molecules with the corresponding number of processors of 64, 128, 256, and 512. These calculations are performed in both KS-DFT (version 6.6 of Quantum ESPRESSO) and eQE 2.0 with both PBE and rVV10 XC functionals. Considering that the number of SCF iterations for different systems is not exactly the same, we report here the wall time for the first 20 iteration steps. As shown in Figure \ref{FNL}, eQE scales ideally when system size and resources are proportionally scaled. This is in contrast with version 6.6 of Quantum ESPRESSO which, as expected, scales polynomially with system size. Most importantly, The computational cost of nonlocal sDFT traces almost perfectly the scaling of semilocal sDFT. These results indicate that the implementation of nonlocal functionals for both NAKEs and NAXC are nearly ideal in eQE 2.0.   

\subsection{Deorbitalized meta GGA functionals}

% Why mGGA?
 A key challenge in DFT is the development of accurate and computationally affordable functionals for the XC energy. The so-called Perdew-Schmidt Jacob's ladder \cite{TPSSH,Perdew_2001} provides a classification of the level of complication that goes in the formulation of the functional. The ladder leads to the ``heaven of chemical accuracy", starting from the local density approximation (LDA; dependence on $n(\br)$, only), generalized gradient approximations (GGA; dependent on $n(\br)$ and $|\nabla n(\br)|$), and meta-generalized gradient approximations (mGGA; dependent on $n(\br)$, $|\nabla n(\br)|$ as well as the noninteracting kinetic energy density, $\tau(\br)$, and possibly the density laplacian, $\nabla^2n(\br)$, and higher order derivatives). Generally, and particularly in this work, the functional dependencies of a mGGA functional can be expressed as follows
\beq
\label{mggaf}
E_{xc}:=E_{xc}\Big[\xi[n,\nabla n],\tau\Big],
\eeq
where the $\xi$ function represents the GGA component of the mGGA functional.

Among all functionals, GGAs dominate materials modeling mainly due to their computational efficiency and generally reliable description of various types of materials \cite{Medvedev_2017}. However, as mentioned, GGAs are severely limited because their description of long/intermediate-range electron-electron correlation, such as \vdw\ interaction, is not satisfactory. Additionally, the presence of self-interaction \cite{Perdew1981} is detrimental for spin-polarized systems and fractional-spin systems (static correlation) \cite{Cohen_2011,cohen2008a}. Among the various mGGA XC functionals, the strongly constrained and appropriately normed (SCAN) satisfies a record 17 exact conditions and can considerably improve GGAs for various systems (e.g., covalent, metallic, and even weak bonds) \cite{Sun_2015}. 

% The issues and challege due to the use of tau
Unfortunately, because $\tau(\br)$ is typically defined in terms of the KS-DFT occupied orbitals $\{\psi_{i}\}$,
\beq
\tau(\br)=\frac{1}{2} \sum_{i=1}^{N_e} \left|\nabla \psi_{i}(\br)\right|^2,
\eeq   
the mGGA XC potential, 
\beq
v_{xc}(\br) =\frac{\delta E_{xc}\Big[\xi[n,\nabla n],\tau\Big]}{\delta n(\br),}
\label{vxc}
\eeq
cannot be directly evaluated by means of functional derivative unless expensive optimized effective potential (OEP) methods are employed \cite{Yang_2002,gorl2005,arbuznikov2003self}.

% The typical approaches to solve the issues in KS-DFT 
To address this complication, a few workarounds have been proposed: (1) completely avoid the evaluation of the $\tau$-dependent mGGA potential. The mGGA functional is just used for non-self-consistent evaluation of the energy using the $n$ and $\tau$ obtained from a non-$\tau$-dependent potential \cite{perdew1999accurate,ernzerhof1999kinetic}. However, this approach is not self-consistent, i.e., the potential is not the derivative of the energy functional which can cause issues, for example, of energy conservation during an ab-initio molecular dynamics. (2) Evaluate the potential in the generalized Kohn-Sham (gKS) scheme. gKS only involves functional derivatives with respect to the KS orbitals \cite{TPSSH,womack2016self} (e.g., $\delta E_{xc}[\{\psi_{i}\}]/\delta \psi_{i}$) yielding an orbital-dependent XC potential. The computational cost of the gKS scheme is much cheaper than OEP, thus it is currently a widely-used approach for $\tau$-dependent mGGAs.  
We note that sDFT cannot directly take advantage of the gKS scheme because only
the KS orbitals of the subsystems are available. The KS orbitals of the global
supersystem are never computed. This issue persists not only for mGGAs, but
also for the Hartree-Fock (HF) exchange potential \cite{lari2010,lari2012}. 

In eQE 2.0, we
decided to approach the problem from a different angle. Inspired by recent
advances in the OF-DFT community, the kinetic energy density, $\tau$, can be
approximated directly with a pure density functional \cite{perdew2007laplacian,mejia2017deorb}. This strategy goes by the name of ``deorbitalization'' converting the explicit orbital-dependent mGGA XC energy
functional to a pure density functional. As a result, the corresponding deorbitalized XC
potential can be evaluated {\it via} \eqn{vxc}
\cite{mejia2017deorb,mejia2018deorbitalized}.

% Tall about the choice of KEDF.
With deorbitalization, the challenge is to approximate the kinetic energy density, not the potential or the energy. Thus, functionals that work well in OF-DFT and as nonadditive functionals in sDFT may in the end not work well for approximating the $\tau$ needed by mGGA XC functionals.  GGAs might be considered, given their easy and cheap evaluation. However, many researchers \cite{lari2014,Smiga_2017,Laricchia_2013b} have shown that Laplacian-level functionals (i.e., those depending on $\nabla^2 n$) are the most suitable functionals for this task.
 
% Details of this approach
%To better show the formalisms of mGGA functionals implemented in eQE,  we first define the following reduced gradient and reduced density Laplacian:
%\begin{enumerate}
%\item Reduced gradient
%\beq
%s=\frac{|\nabla n(\br)|}{2(3\pi^2)^{1/3}n^{4/3}(\br)}
%\label{s}
%\eeq
%\beq
%p=s^2 = C_{p}\frac{|\nabla n(\br)|^2}{ n^{8/3}(\br)}
%\label{p}
%\eeq
%\item Reduced density Laplacian
%\beq
 %q= \frac{\nabla^2n(\br)}{4(3\pi^2)^{2/3}n^{5/3}(\br)}= C_{p}\frac{\nabla^2 n(\br)}{n^{5/3}(\br)}
%\label{q}
%\eeq
%\end{enumerate}
%where $C_p =\frac{1}{4(3\pi^2)^{2/3}}$.
 Laplacian-level KEDF approximations can be written as follows
%\beq
%T_{s}[n,\nabla n,\nabla^2n] =\int d \br \tau_{\mathrm{TF}}(\br) F_{t}[p,q](\br)=\int \tau[n,\nabla n, \nabla^{2}n](\br) d\br.
%\label{tau}
%\eeq
\beq
T_{s}[n,\nabla n,\nabla^2n] =\int \tau[n,\nabla n, \nabla^{2}n](\br) d\br.
\label{tau}
\eeq
%Where the kinetic energy density  $\tau = \tau_{\mathrm{TF}}(\br)F_{t}[p,q](\br) ,$ $ \tau_{\mathrm{TF}} =C_{\mathrm{TF}}n^{5/3}(\br)$, and $C_{\mathrm{TF}} =\frac{3}{10}(3\pi^2)^{2/3}$.
Thus, the deorbitalized mGGA functionals can be written as,
\beq
E_{xc} =\int \varepsilon_{xc}\Big[\xi[n,\nabla n],\tau[n,\nabla n,\nabla^{2}n]\Big](\br) d\br=\int \varepsilon_{xc} [n,\nabla n,\nabla^{2}n](\br) d \br.
\label{fexc}
\eeq

In this way, the XC functional is an explicit density functional and the corresponding XC potential can be evaluated by \eqn{vxc}. Namely,
\beq
v_{xc}=\frac{\delta E_{xc}}{\delta n} =\frac{\partial \varepsilon_{xc}}{\partial n} -\nabla \cdot(\frac{\partial \varepsilon_{xc}}{\partial \nabla n}) + \nabla^{2}(\frac{\partial \varepsilon_{xc}}{\partial \nabla^{2} n}).
\label{fvxc}
\eeq
where we are omitting hereafter specific variable dependence of the density and other functions/functionals. Considering $\tau$ as a functional of $n$, $\nabla n$, and $\nabla^{2}n$ as shown in \eqs{tau}{fexc}, the XC energy density partial derivatives with respect to $n$, $\nabla n$, and $\nabla^{2}n$ can be expressed as follows
\beq 
\frac{\partial \varepsilon_{xc}}{\partial n} ={ \color{red} \frac{\partial \varepsilon_{xc}}{\partial \xi} \frac{\partial \xi}{\partial n} }+ {\color {red}\frac{\partial \varepsilon_{xc}}{\partial \tau}}\frac{\partial \tau}{\partial n}
\eeq

\beq
\frac{\partial \varepsilon_{xc}}{\partial \nabla n} ={\color {red} \frac{\partial \varepsilon_{xc}}{\partial \xi} \frac{\partial \xi}{\partial \nabla n}} + {\color {red} \frac{\partial \varepsilon_{xc}}{\partial \tau}}\frac{\partial \tau}{\partial \nabla n}
\eeq

\beq
\frac{\partial \varepsilon_{xc}}{\partial \nabla^2 n} = {\color{red} \frac{\partial \varepsilon_{xc}}{\partial \tau}}\frac{\partial \tau}{\partial \nabla^2 n}.
\eeq
Here, the terms in red are independent of the choice of $\tau$. Thus, these operations are collected in a single subroutine for each mGGA XC functional.

To better show the formalisms of mGGA functionals related to $\tau$ parts implemented in eQE,  we first define the following reduced gradient and reduced density Laplacian:
\begin{enumerate}
\item Reduced gradient
\beq
s=\frac{|\nabla n|}{2(3\pi^2)^{1/3}n^{4/3}}
\label{s}
\eeq
\beq
p=s^2 = C_{p}\frac{|\nabla n|^2}{ n^{8/3}}
\label{p}
\eeq
\item Reduced density Laplacian
\beq
 q= \frac{\nabla^2n}{4(3\pi^2)^{2/3}n^{5/3}}= C_{p}\frac{\nabla^2 n}{n^{5/3}}
\label{q}
\eeq
\end{enumerate}
where $C_p =\frac{1}{4(3\pi^2)^{2/3}}$.

In general, Laplacian-level kinetic energy density $\tau$ can be expressed as :
\beq
\tau = \tau_{\mathrm{TF}}F_{t}[p,q],
\label{ftau}
\eeq
where  $ \tau_{\mathrm{TF}} =\frac{3}{10}(3\pi^2)^{2/3}n^{5/3}$.

 Using of \eqn{ftau}, the kinetic energy density partial derivatives with respect to $n$, $\nabla n$, and $\nabla^{2}n$ can be further written as follows
\beq
\label{tau1}
\frac{\partial \tau}{\partial n} =\left( {\color {blue} \frac{\partial F_{t}}{\partial p}}\frac{\partial p}{\partial n}+ {\color{blue} \frac{\partial F_{t}}{\partial q}}\frac{\partial q}{\partial n}\right)\tau_{\mathrm{TF}} + \frac{\partial \tau_{TF}}{\partial n}{\color{blue}F_{t}}
\eeq
\beq
\label{tau2}
\frac{\partial \tau}{\partial \nabla n} ={\color {blue} \frac{\partial F_{t}}{\partial p}}\frac{\partial p}{\partial \nabla n}\tau_{\mathrm{TF}}
\eeq

\beq
\label{tau3}
\frac{\partial \tau}{\partial \nabla^2 n} ={\color {blue} \frac{\partial F_{t}}{\partial q}}\frac{\partial q}{\partial \nabla^2 n}\tau_{\mathrm{TF}},
\eeq
where 
$\frac{\partial p}{\partial n} =-\frac{8}{3}p/n$, $\frac{\partial q}{\partial n} =-\frac{5}{3}q/n$, $\frac{\partial p}{\partial \nabla n} /\nabla n=\frac{2C_p}{n^{8/3}}$, $\frac{\partial q}{\partial \nabla^2 n} =\frac{C_p}{n^{5/3}}$, and $\frac{\partial \tau_{TF}}{\partial  n} = \frac{5}{3}C_{\mathrm{TF}} n^{2/3}$.
The above equations (\ref{tau1}--\ref{tau3}) are independent of the particular mGGA XC functional used. The terms in blue color are determined only by the choice of KEDF, and the other terms are independent of the choice of both XC and KEDF. Thus, the terms in blue are evaluated in a separate, independent subroutine which is called by the subroutine that combines all terms together.
%In evaluation of the $E_{\rm xc}$ (Eq.\ref{fexc}) and the corresponding effective XC potential $v_{xc}(\br)$ (Eq.\ref{fvxc}), the choice of mGGA XC functionals and the KEDF used for evaluation of kinetic energy density $\tau$ are independent in this formalism. Thus, in eQE framework the calculators for the KEDF ( i.e.  $F_t$,  $\frac{\partial F_{t}}{\partial p}$,  and $\frac{\partial F_{t}}{\partial q}$) are evaluated in one subroutine, and the calculators just related the mGGA functional itself (i.e. $\frac{\partial \tau_{xc}}{\partial n}$, $\frac{\partial \tau_{xc}}{\partial \nabla n}$, $\frac{\partial \tau_{xc}}{\partial \tau}$) are evaluated in another subroutine. 

% what we really have done
eQE 2.0 implements one of the best performing mGGA functionals, the SCAN functional \cite{Sun_2015} deorbitalized utilizing Laplacian-level KEDFs (PC \cite{perdew2007laplacian}, PCopt \cite{mejia2017deorb}, TFLopt \cite{mejia2017deorb}, GEA2L, L04 \cite{lari2014}). 

% additional techniques to solve the numerical problem
\subsection{A note on numerical stability}
Previous works \cite{mejia2017deorb,mejia2018deorbitalized} show that the use of the Laplacian operator can introduce noise in the results, especially when a plane-wave basis is adopted. To address this challenge we implemented a smooth Laplacian, $\hat{L_s}$, inspired by convolutions typically used for smoothing images. Consider $x(\br)$ as any physical quantity defined in real space (such as electron density $n(\br)$ and potential $v(\br)$), then, the smooth Laplacian operator, $\hat{L_s}$, operates on $x(\br)$ as,
\beq
\hat{L_s} x(\br) = -\F^{-1}\left[\F\left[\hat{S}\left[x(\br)\right] +\hat{S}\left[x(\br)-\hat{S}[x(\br)]\right]\right]|\mathbf{g}|^2e^{-\sigma |\mathbf g|^2}\right].
\eeq
Where $\hat{S}$ is defined as, 
\beq
\hat{S} x(\br) =\F^{-1}\left[\F[x(\br)]e^{-\sigma |\mathbf g|^2}\right],
\eeq 
$\mathbf g$ is the reciprocal space vector, and $\sigma$ is a parameter. Based on our tests, $\sigma=0.02$ is a good choice to maintain numerical stability without over smoothing.

%Benchmarks
To verify the SCAN functional implemented in eQE, we implemented the subroutines we use in eQE for evaluation of SCAN functional in Quantum ESPRESSO 6.6.  Our implementation of SCAN in Quantum ESPRESSO (using the gKS approach) reproduced exactly the same results as the latest version Quantum ESPRESSO with Libxc \cite{lehtola2018recent,marques2012libxc}. This indicates that our implementation is correct. As previous work shows \cite{lari2014}, the L04 mGGA KEDF outperforms other mGGA KEDFs when employed to approximate $\tau$ for mGGA XC functionals. Thus, here, we select L04 in all calculations. We note, however, that the PCopt KEDF has also been reported to be suitable for this task \cite{mejia2018deorbitalized}.
 \begin{figure}
\includegraphics[width=1.0\textwidth]{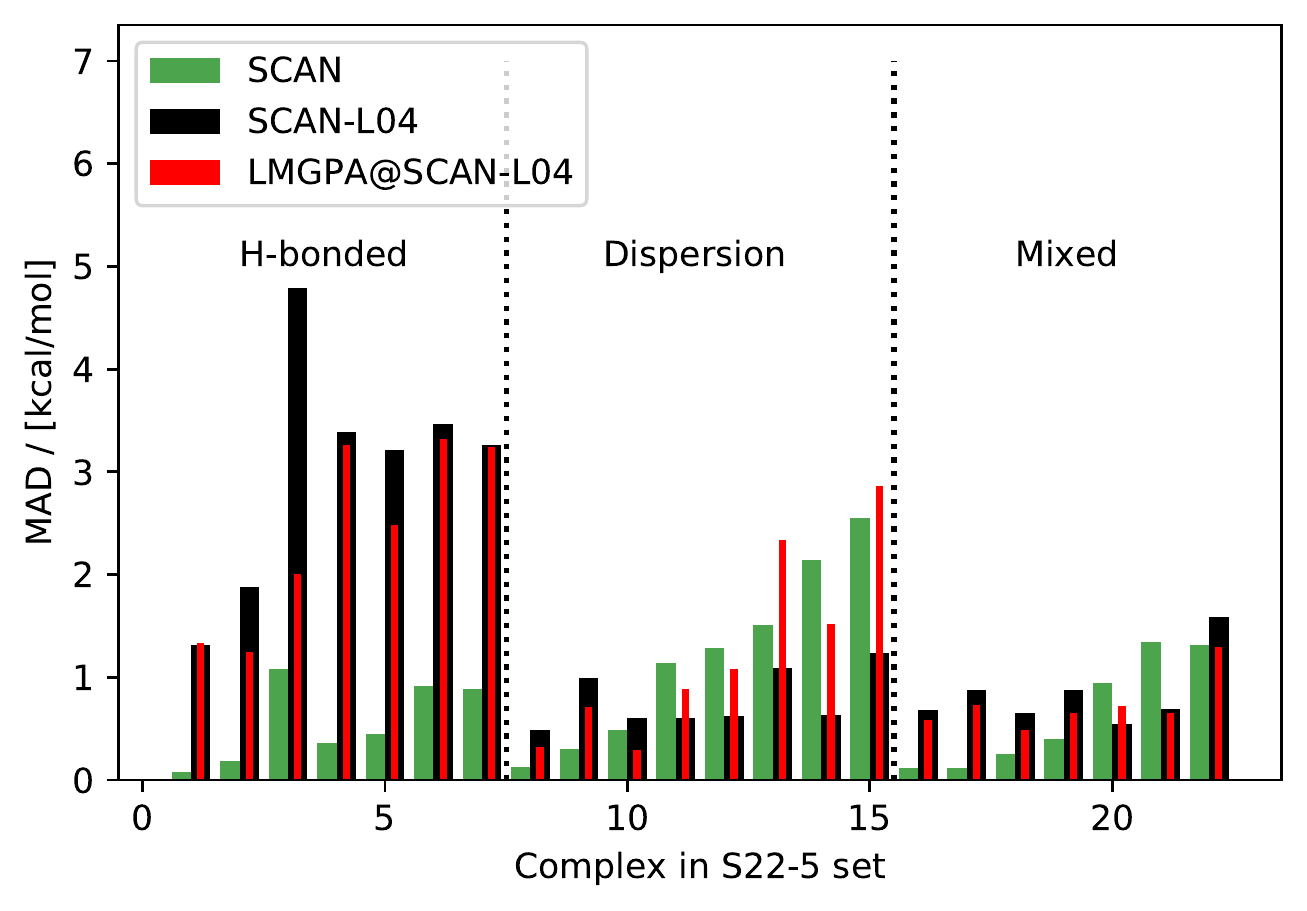}
\caption{\label{mGGA} Mean absolute deviations (MADs) of interaction energies
computed by several realizations of the SCAN functional: original (SCAN), deorbitalized SCAN with L04 KEDF for $\tau$ (SCAN-L04), and SCAN--L04 in sDFT coupled with nonlocal NAKE (LMGPA@SCAN-L04) in comparison
to CCSD(T) for the S22-5 test set. The SCAN results are taken from Ref.\ \citenum{Sun_2015}. All values are given in kcal/mol. H-Bond, Dispersion and Mixed stand for hydrogen-bonded, dispersion-bonded and complexes bonded by a mix of hydrogen bonds and dispersion interactions, respectively.}
\end{figure}
 
\begin{table*}[h!]
\centering
\caption{\label{tab:mGGA}Summary of the Mean absolute deviations (MADs) of interaction energies computed with different approaches against CCSD(T) results. All values are given in kcal/mol. H-bonded, Dispersion and Mixed stand for hydrogen-bonded, dispersion-bonded and complexes bonded by a mix of hydrogen bonds and dispersion interactions, respectively.}
\begin{tabular}{c c c c c c}
\hline
\hline
Methods                        & H-bonded & Dispersion & Mixed & Total \\
\hline
SCAN                           & 0.56 & 1.19 & 0.64 & 0.82 \\
SCAN-L04                    & 3.04 & 0.79 & 0.84 & 1.52 \\
LMGPA@SCAN-L04    & 2.41 & 1.25 & 0.73 & 1.46  \\
\hline
\hline
\end{tabular}
\label{tab: test_system}
\end{table*}
We again select the S22-5 test set with the same settings as before\cite{Mi_2019}, except for the plane wave cutoffs which is increased to 500 Ry for the density and is kept to 70 Ry for the wavefunctions. As shown in Figure \ref{mGGA}, we present the mean absolute deviations (MADs) of interaction energies obtained with different approaches in comparison to the benchmark CCSD(T) energies. These approaches include the SCAN functional implemented in the gKS scheme (SCAN), the SCAN-L04 functional (SCAN-L04), and the SCAN-L04 functional used in sDFT with the nonlocal LMGPA as NAKE (LMGPA@SCAN-L04). It is clear that SCAN-L04 delivers good interaction energies with MADs well below 5 kcal/mol. SCAN-L04 does not deteriorate the good performance of SCAN, except for the hydrogen bonded systems. Moreover, sDFT simulations with LMGPA@SCAN-L04 combination can deliver no worse results (even better for hydrogen bonded and Mixed bonded system) in comparison to SCAN-L04 results.       

To further quantify the performance of SCAN-L04, we summarize the MADs of the interaction energies calculated with SCAN, SCAN-L04, and LMGPA@SCAN-L04 against CCSD(T) results in Table \ref{tab:mGGA}. It is clear that the performance of these three approaches are similar, since the total MAD for
SCAN, SCAN-L04, and LMGPA@SCAN-L04 are 0.82, 1.52, 1.46 kcal/mol, respectively. For hydrogen-bonded systems, SCAN (0.56 kcal/mol) outperforms LMGPA@SCAN-L04 (2.41 kcal/mol) and SCAN-L04 (3.04 kcal/mol).   

% Issues 
It is clear that deorbitalized mGGA functionals are competitive against the best pure functionals currently on the market. We envision further improvement for the deorbitalization strategy tackling the following weaknesses:
\begin{enumerate}
\item Deorbitalized mGGAs require larger plane-wave cutoffs to achieve converged results compared to typical GGA functionals. 
\item The use of Laplacian-level KEDF for evaluating $\tau$ with plane-wave basis sets is numerically challenging. This can lead to an increase in the number of SCF iterations.
\item Currently available approximations to $\tau[n](\br)$ are still fairly crude.
\end{enumerate}

\section{Conclusions and future directions}
\label{sec:conclusion}
This work releases version 2 of embedded Quantum ESPRESSO -- a code for running density embedding simulations of molecules and condensed-phase systems.  Since its first release, eQE has been successfully employed for many large-scale simulations for condensed matter physics and chemistry. The new release allows the user to employ nonlocal functionals both for the kinetic energy (needed for density embedding) and exchange-correlation.  Specifically, the new functionals include nonlocal nonadditive kinetic energy such as LMGPA, and nonlocal XC functionals such as rVV10. eQE version 2 also includes deorbitalized meta GGA XC functionals (i.e., where the kinetic energy density, $\tau$, is approximated by a pure density functional). Preliminary benchmark tests for the new nonlocal functionals indicate that they considerably improve on the performance of semilocal functionals in KS-DFT and subsystem DFT (embedding) calculations without increasing the computational cost significantly and, most importantly, maintaining eQE's strong scalability with systems size. 

Following this release, we will turn to further improve eQE's performance implementing new features which may include (but not limited to) development of new nonadditive XC functionals, especially the latest SCAN variants (such as: r2SCAN \cite{furness2020accurate,mejia2020meta}, r2SCAN+ \cite{ehlert2020r2scan,grimme2020r2scan}, SCAN+rVV10 \cite{peng2016versatile}). We will also continue to develop more accurate KEDFs for both nonadditive functionals for embedding and for approximating the  kinetic energy density used in deorbitalized meta GGA functionals. 

\section{Acknowledgments} 
This material is based upon work supported by the
National Science Foundation under Grants No.\ CHE-1553993 and OAC-1931473. We thank the Office of Advanced Research Computing at Rutgers for providing access to the Amarel and Caliburn clusters. 
%\subsection
\bibliographystyle{elsarticle-num}
\bibliography{../prg_bibliography/prg.bib}

\end{document}